\documentstyle[twocolumn,aps]{revtex}

\begin{document}

\title{Exact density of states of a two-dimensional electron gas
in a strong magnetic field and a long-range correlated
random potential.}

\author{Lothar Spies$^1$, Walter Apel$^2$ and Bernhard Kramer$^1$}

\address{$^1$I. Institut f\"ur Theoretische Physik, Universit\"at
Hamburg, Jungiusstrasse~9, D-20355 Hamburg, Germany\\
$^2$Physikalisch-Technische Bundesanstalt, Bundesallee 100, D-38116
Braunschweig, Germany
\smallskip\\
{\rm (September 30, 1996)}
\bigskip\\[0.5cm]
\parbox{14.2cm}{\rm
We derive an exact result for the averaged Feynman propagator and the
corresponding density of states of an electron in two dimensions in a
perpendicular homogeneous magnetic field and a Gaussian random potential
with long-range spatial correlations described by a
{\em quadratic correlation function}.
}}

\maketitle
\narrowtext
\vspace{1cm}

The problem of an electron moving in two dimensions (2D) in a strong
homogeneous magnetic field $B$ and a random potential has been
studied particularly intensively since the discovery of the integer
quantum Hall effect (IQHE) \cite{kdp80}. Especially the density of
states (DOS) as the simplest quantity that characterizes the quantum
mechanical spectrum has been the subject of continuous interest
\cite{afs82,w83,bgi84,skm84,gg86,a87,scg88,bnl89,bhkl95}. This was
partly induced by the fact that it remains notoriously difficult
to explain the unexpectedly large density of states between the
Landau bands (LBs) detected experimentally
\cite{lg83,eetal85,getal85,hetal86,metal86,as92,petal96}. The
theoretical calculations, mostly based on models with short-range
correlated randomness as, for instance, the Gaussian white noise
potential, yielded in the high magnetic field limit well separated,
disorder broadened LBs with a width proportional to $\sqrt{B}$ and an
exponentially small DOS in between the bands \cite{g75,au74}. On the
other hand, the experimental data indicate that there is a roughly
constant background DOS which contains as many as 10\% to 20\% of the
levels \cite{lg83}.

It is very tempting to suspect that spatially long-range correlations
of the disorder, due to inhomogeneities, for instance, are responsible
for this background DOS. Such a model was studied by using path
integrals in connection with a cumulant expansion \cite{scg88}. This
reproduced the experimental result. However, the influence of the
cumulant approximation which is uncontrolled in the presence
of a magnetic field was unclear. Subsequent approximative treatments of
the DOS in the presence of long-range correlated disorder, by using
functional integrals and without the cumulant approximation, did not
improve the situation \cite{bnl89}. Exact results for long-range
correlated disorder exist only for the band tails \cite{a87}.

In this paper, we consider a model that has already been treated before
without magnetic field \cite{b70}. Also in the region of the quantum
Hall effect, this model allows for the treatment of the influence of
long-range correlations in the randomness.  The essential assumption
is a quadratic approximation of the correlation function. The range of
validity of this model assumption is discussed below. The Feynman
propagator can be calculated exactly, and the DOS evaluated
numerically, but without employing further approximations.

The comparison with recent experimental results \cite{h96} shows that
long-range correlations can qualitatively explain some aspects of the
experimental data. However, the quantitative behavior, as a function
of the magnetic field, cannot be understood. We conclude, that the
large DOS between the LBs seen in experiment, is very likely related
to interaction between the electrons.

We consider a spinless electron moving in 2D under the  influence of a
homogeneous magnetic field and a random impurity potential
$V(\mbox{\boldmath$r$\unboldmath})$.  The Hamiltonian is
\begin{equation}
H = \frac{1}{2}
        \left[ i\mbox{\boldmath$\nabla$\unboldmath} +
        \mbox{\boldmath$A$\unboldmath}
                (\mbox{\boldmath$r$\unboldmath})\right]^2 +
        V(\mbox{\boldmath$r$\unboldmath}).
\label{landau}
\end{equation}
For the vector potential, we assume the symmetric gauge
$\mbox{\boldmath$A$\unboldmath}(\mbox{\boldmath$r$\unboldmath}) =
\hat{\epsilon} \cdot\mbox{\boldmath$r$\unboldmath} /2 $,
where $\hat{\epsilon}$  denotes the antisymmetric $(2\times 2)$
matrix, so that $\hat{\epsilon}^2 = -1$.
The magnetic field $B$ defines the
intrinsic energy, length, and time scales. Energies and lengths are
measured here in units of the  cyclotron energy $\hbar\omega _{c}\equiv
\hbar eB/m$ and the magnetic length $\ell _{c}\equiv \sqrt{\hbar/eB}$,
respectively ($e$ elementary charge and $m$ effective mass). The unit 
of time is then the inverse of $\omega _{c}$.
The probability distribution of the impurity potential
$V(\mbox{\boldmath$r$\unboldmath})$
is assumed to be Gaussian with zero mean
$\langle V(\mbox{\boldmath$r$\unboldmath}) \rangle = 0$
and correlation function
\begin{equation}
C(\mbox{\boldmath$r$\unboldmath},\mbox{\boldmath$r$\unboldmath}')\equiv
\langle V(\mbox{\boldmath$r$\unboldmath})
V(\mbox{\boldmath$r$\unboldmath}')\rangle = \eta^2 \left[ 1 -
\frac{1}{L^2}\left(\mbox{\boldmath$r$\unboldmath} -
\mbox{\boldmath$r$\unboldmath}'\right)^2\right].
\label{correlationfunction}
\end{equation}
Angular brackets denote the ensemble average with respect to the
probability distribution of the random variable $V$.
The parameters $\eta$ and $L$ determine the strength and the range
of the impurity potential, respectively.

The above polynomial correlation function Eq.~(\ref{correlationfunction})
is the first order expansion of a Gaussian
\begin{equation}
C_{G}(\mbox{\boldmath$r$\unboldmath},\mbox{\boldmath$r$\unboldmath}')=
\eta ^{2}e^{-\left(\mbox{\boldmath$r$\unboldmath} -
\mbox{\boldmath$r$\unboldmath}'\right)^2/L^{2}}.
\label{gaussian}
\end{equation}
The function
$C(\mbox{\boldmath$r$\unboldmath},\mbox{\boldmath$r$\unboldmath}')$,
Eq.~(\ref{correlationfunction}), has been used previously \cite{b70}
to describe approximately the DOS of an electron in 2D in a random
impurity potential with the Gaussian correlation function,
Eq.~(\ref{gaussian}).

Here, we derive the exact result for the impurity averaged Feynman
propagator of the model Eq.~(\ref{correlationfunction}),
\begin{equation}
\langle K(\mbox{\boldmath$r$\unboldmath},t;
                \mbox{\boldmath$r$\unboldmath}_0)\rangle
= \langle \int\limits
                _{\mbox{\boldmath$r$\unboldmath}(0)
                =\mbox{\boldmath$r$\unboldmath}_0 }
                ^{\mbox{\boldmath$r$\unboldmath}(t)
                =\mbox{\boldmath$r$\unboldmath}}
{\cal D}[\mbox{\boldmath$r$\unboldmath}]
\exp{\left[i\int_{0}^{t}d\tau {\cal L}(\mbox{\boldmath$r$\unboldmath},
                \dot{\mbox{\boldmath$r$\unboldmath}};\tau)\right]}\rangle ,
\label{propagator}
\end{equation}
where ${\cal D}[\mbox{\boldmath$r$\unboldmath}]$ is the Feynman
measure on the set of trajectories with boundary conditions
$\mbox{\boldmath$r$\unboldmath}(0)= \mbox{\boldmath$r$\unboldmath}_0$
and $\mbox{\boldmath$r$\unboldmath}(t)=
\mbox{\boldmath$r$\unboldmath}$. The time evolution begins at $t=0$,
which is omitted in
$K(\mbox{\boldmath$r$\unboldmath},t;\mbox{\boldmath$r$\unboldmath}_0)$
for simplicity. Equation~(\ref{propagator}) is a Feynman path integral
and the Langrangian
\begin{equation}
{\cal L}(\mbox{\boldmath$r$\unboldmath},
        \dot{\mbox{\boldmath$r$\unboldmath}};\tau) =
        \frac{1}{2}\dot{\mbox{\boldmath$r$\unboldmath}}^{2} +
        \frac{1}{2}\dot{\mbox{\boldmath$r$\unboldmath}}^{T}
        \hat{\epsilon}\mbox{\boldmath$r$\unboldmath} -
        V(\mbox{\boldmath$r$\unboldmath})
\label{lagrangian}
\end{equation}
represents the usual expression for a charged particle in 2D
in a magnetic field and a potential.

Since $V(\mbox{\boldmath$r$\unboldmath})$ enters only linearly in
the expo\-nent of Eq.~(\ref{propagator}), the
impurity average is ea\-si\-ly per\-for\-med,
\begin{eqnarray}
&&\langle K(\mbox{\boldmath$r$\unboldmath},t;
\mbox{\boldmath$r$\unboldmath}_0)\rangle =
\int\limits_{\mbox{\boldmath$r$\unboldmath}(0)=
\mbox{\boldmath$r$\unboldmath}_0 }^{\mbox{\boldmath$r$\unboldmath}(t)=
\mbox{\boldmath$r$\unboldmath}}
{\cal D} [\mbox{\boldmath$r$\unboldmath}] \exp{\left\{\int_0^t d\tau
\frac{i}{2}\left[\dot{\mbox{\boldmath$r$\unboldmath}}^{2} +
\dot{\mbox{\boldmath$r$\unboldmath}}^{T}
\hat{\epsilon}\mbox{\boldmath$r$\unboldmath}\right]\right\}} \nonumber\\
&&\nonumber\\
&&\qquad\quad\times
\exp \left\{ -\frac{1}{2} \int_0^t d\tau \int_0^t d\tau' \;
C(\mbox{\boldmath$r$\unboldmath}(\tau ),
                \mbox{\boldmath$r$\unboldmath}(\tau '))\right\}.
\label{pathint}
\end{eqnarray}
The last factor in Eq.~(\ref{pathint}) results from the impurity average.
It is purely real and non-local in time. This consequence of integrating
out degrees of freedom is well known \cite{khandekar}.
It is usually the end of an exact calculation.
In our model, Eq.~(\ref{correlationfunction}), however, we have the
simplifying feature that
$C\left(\mbox{\boldmath$r$\unboldmath}, \mbox{\boldmath$r$\unboldmath}'\right)$
is quadratic in
$\mbox{\boldmath$r$\unboldmath} - \mbox{\boldmath$r$\unboldmath}'$.
Therefore, the path integral remains quadratic and can be evaluated
exactly by using standard techniques \cite{papa}
-- in spite of the non-locality in time. Details are given in
\cite{spies}.

Denoting
$\langle K(\mbox{\boldmath$r$\unboldmath},t;\mbox{\boldmath$r$\unboldmath}_0)
\rangle \equiv
K_B(\mbox{\boldmath$r$\unboldmath},t;\mbox{\boldmath$r$\unboldmath}_0)
$, we obtain
\begin{eqnarray}
&&K_B(\mbox{\boldmath$r$\unboldmath},t;\mbox{\boldmath$r$\unboldmath}_0)=
\nonumber\\
&&\nonumber\\
&&\qquad \frac{1}{2 \pi i t} \cdot
\frac{(t^2/4)(\Omega ^2 - 1/4)}
{\sin \left[\left(\Omega + 1/2\right)(t/2)\right]
\cdot \sin \left[\left(\Omega - 1/2\right)(t/2)\right]}
\nonumber  \\
&&\nonumber  \\
&&\qquad\qquad\qquad
\times \exp{\left[-\frac{1}{2} \eta^2 t^2 +i f(\mbox{\boldmath$r$\unboldmath},
\mbox{\boldmath$r$\unboldmath}_{0};t)\right]}
\end{eqnarray}
and
\begin{eqnarray}
f(\mbox{\boldmath$r$\unboldmath},\mbox{\boldmath$r$\unboldmath}_{0};t)
&=& \frac{\Omega }{2 \sin \Omega t}\left[
\frac{\left(\mbox{\boldmath$r$\unboldmath} -
\mbox{\boldmath$r$\unboldmath}_0 \right)^2}{2} \cos \Omega t \right.
\nonumber \\
&&\nonumber \\
&& -2 \mbox{\boldmath$r$\unboldmath}_0^{T} e^{
\frac{1}{2}\hat{\epsilon}t}
\mbox{\boldmath$r$\unboldmath}+\frac{\left(\mbox{\boldmath$r$\unboldmath} +
\mbox{\boldmath$r$\unboldmath}_0 \right)^2}{2}
\cos \frac{1}{2}t \nonumber \\
&&\nonumber\\
&& +\frac{\left(\mbox{\boldmath$r$\unboldmath}-
\mbox{\boldmath$r$\unboldmath}_0 \right)^2}{2 \Omega ^{2}}
\frac{\left(
\Omega \sin \frac{1}{2}t - \frac{1}{2} \sin \Omega
t \right)^2}{\cos \frac{1}{2}t -
\cos \Omega t} \nonumber \\
&&\nonumber \\
&& +\left.\frac{2\mbox{\boldmath$r$\unboldmath}_0^{T} \hat{\epsilon}
\mbox{\boldmath$r$\unboldmath}}{\Omega }\left(\Omega\sin
\frac{1}{2}t - \frac{1}{2}\sin \Omega t\right)
\right].
\end{eqnarray}
The complex frequency
$ \Omega = \frac{1}{2}\sqrt{1+8i t (\eta /L)^2}$
depends on the time $t$ of propagation.
Configurational averaging leads to a path integral with a non-local action.
Therefore, the above is not a conventional Feynman propagator.
It does not have the group property,
\begin{equation}
K_B(\mbox{\boldmath$r$\unboldmath},t;\mbox{\boldmath$r$\unboldmath}_0)
\neq \int d^2 r'
K_B(\mbox{\boldmath$r$\unboldmath},t-t';\mbox{\boldmath$r$\unboldmath}') \,
K_B(\mbox{\boldmath$r$\unboldmath}', t';\mbox{\boldmath$r$\unboldmath}_0),
\end{equation}
which would be valid for a unitary time evolution operator {\em without}
the configurational average.

This has serious consequences for the analytical structure of the diagonal part
\begin{eqnarray}
&&K_B(t)  \equiv
K_B(\mbox{\boldmath$r$\unboldmath},t;\mbox{\boldmath$r$\unboldmath})
=\frac{e^{-(\eta t)^2/2 }}{2 \pi i t}
\nonumber\\
&&\nonumber\\
& &\qquad
\times\frac{(t^{2}/4)\left(\Omega ^{2} - 1/4\right)}
     {\sin \left[\left(\Omega + 1/2\right)(t/2)\right]
\,\sin \left[\left(\Omega - 1/2\right)(t/2)\right]}
\label{diagonalpart}
\end{eqnarray}
of our propagator that determines the DOS. By investigating the
analytical continuation of $K_B(t)$ into the complex $t$-plane, we
find not only a single pole for $t=0$ on the real time axis, but
also poles in the lower and upper complex plane. In the limit $L \to
\infty$, the poles in the lower plane approach the real time axis
whereas the poles in the upper plane  disappear towards infinity.

Evaluating the DOS for such a propagator, we have to take two
conditions into account: first, causality should be fulfilled and
second, the definition of the DOS for this model should reproduce
the limit $L=\infty$ of a model with constant spatial correlations.
The DOS can be calculated exactly in this limit \cite{edw,lukes}.
We choose the integration contour in the complex $t$-plane
in such a way that coming from minus infinity all poles, $t_n$,
in the lower half plane are passed anti-clockwise. This choice is
unique, and the limit $L\to \infty$ is continuous.
Deforming this contour back to the real axis gives the DOS
\begin{eqnarray}
&&D(E) = \frac{1}{4\pi}\quad +\quad \frac{1}{2\pi }
{\cal P}\int\limits_{-\infty}^{\infty} dt
K_B(t)e^{iEt}
\nonumber\\
&& \qquad\qquad + i\sum_{\mbox{\footnotesize{Im}}(t_{n}) <0}
\mbox{Res}_{t=t_n}\left[ K_B(t) e^{iEt} \right].
\label{result}
\end{eqnarray}
Here, ${\cal P} \int_{-\infty}^{\infty} dt \ldots \;$ denotes
Cauchy's principal value of the integral
and $\mbox{Res}_{t=t_n}\left[\ldots \right]$ the residue at $t=t_n$.

Equation~(\ref{result}) is our central result.
Figure~1 shows the result of the integration for $L=6$ and $\eta
=0.2$, $0.3$ and $0.4$ (in the units defined above), normalized to the DOS
at $B = 0$, $D_0$.
Due to the impurity potential, the Landau peaks are broadened as expected.
The broadening decreases for increasing
energy which indicates that disorder becomes less effective.
For large $L$, Eq.~(\ref{result}) can be shown to yield the same
as obtained for the model with a constant correlation function,
a superposition of Gaussians centered at the Landau levels (LLs).
How\-ever, Eq.~(\ref{result}) becomes negative if $E$ exceeds a
critical value.
This is a consequence of the unphysical shape of the polynomial
correlation function Eq.~(\ref{correlationfunction}), as we will now
see.

While the correlation function Eq.~(\ref{gaussian}) is positive
definite, its truncated expansion Eq.~(\ref{correlationfunction}) is
not. Nevertheless, using Eq.~(\ref{correlationfunction}) in calculating
physical quantities may give the same result as using
Eq.~(\ref{gaussian}) and expanding afterwards in $L^{-1}$.
It is instructive to take the limit in Eq.~(\ref{diagonalpart})
in which one keeps only the lowest LL in Eq.~(\ref{landau}).
Formally, this is achieved by measuring the energy from the center of
the lowest LL, scaling $\tilde{E} = \hbar\omega _{c} (E - 1/2)$ and
$\tilde{\eta } = \hbar\omega _{c} \eta $, and taking
$\omega _{c}\to \infty$.
We find for the DOS of the lowest LL with the random potential
described by Eq.~(\ref{correlationfunction}),
\begin{equation}
D_{LLL}(\tilde{E})=\frac{1}{2\pi}
\int\limits_{-\infty}^{\infty}\frac{dt}{2\pi}
e^{-(\tilde{\eta }t)^{2}/2}\;
\frac{ 2(\tilde{\eta }t/L)^{2} \; \cos \tilde{E}t }
{1-\exp{(-2(\tilde{\eta }t/L)^{2})}}.
\end{equation}
The limit $L\to\infty$ is correctly reproduced. For $L>\sqrt{2}$,
$D_{LLL}(\tilde{E})$ is positive for all energies. Now, the
following argument suggests that including more and more LLs
yields an increasingly higher bound for $L$ until it diverges in the
limit of all LLs taken into account. The classical cyclotron radius
is proportional to $\sqrt{E}$. Thus, the correlation function
Eq.~(\ref{correlationfunction}) is positive and the DOS well defined only
for energies smaller than $E_{\mbox{{\small max}}}\approx (L/2)^{2}$,
independent of the strength of the disorder $\eta $. Beyond this
critical energy, the classical cyclotron radius, which is a
characteristic length for the range of the electron, is of the order of
$L$ and larger.
Then, the expansion leading from Eq.~(\ref{gaussian}) to
Eq.~(\ref{correlationfunction}) breaks down and, for such distances,
the correlation function Eq.~(\ref{correlationfunction}) becomes negative.
Such a bound $E_{\mbox{\small max}}$, and its independence of $\eta$,
can indeed be seen in Fig.~1.

Our model is the first order approximation of an impurity potential with
Gaussian correlations, Eq.~(\ref{gaussian}), which is believed
to be typical for such materials as GaAs/AlGaAs heterostructures
\cite{afs82}. In order to obtain information on the validity of our
model, Eq.~(\ref{correlationfunction}), we compare the
DOS with previous results for the Gaussian model.

By neglecting the mixing of states between different LLs and
assuming a Gaussian shape of each level, Broderix {\it et al.} \cite{brod}
found for the  width of the $n^{\mbox{\small th}}$ LB
\begin{equation}
\sigma_n^2 = \eta ^2\frac{L^2}{L^2+2}\left[\frac{L^2-2}{L ^2+2}\right]^n
P_n\left(\frac{L^4+4}{L ^4-4}\right),
\end{equation}
with $P_n(x)$ the $n^{\mbox{\small th}}$ Legendre polynomial.
Figure~2 shows both results, ours and that of \cite{brod}, for $L=10$ and
$\eta = 0.4$. The agreement is quantitative. For increasing $L$,
the agreement becomes even better.
This demonstrates indeed that for large values, the length
parameter $L$ has the meaning of the correlation length of a Gaussian
correlation function. Also, our results support the conjecture that in
the quantum Hall regime, LL mixing is negligible and the DOS
is a superposition of Landau bands with Gaussian shapes.
This is valid for even a broader energy range when $L$ is increased.

Armed with the exact result for the DOS of a model with spatially
correlated randomness, we now compare with experiments, cf.~Fig~3. By
plotting the DOS at the cyclotron energy  -- between the lowest and the
second lowest LL -- against the magnetic field strength, we find
that the various  experiments on GaAs/AlGaAs heterostructures show the
same qualitative behavior. However, this  appears to be quite different
from the prediction based on the model of non-interacting
electrons with spatially correlated disorder. We find a strong
decrease of the DOS when increasing the strength of the magnetic
field, much stronger than is observed experimentally. Since a
spatially correlated random potential energy represents the most
general model of disorder, this leads to the question, whether or not
the interaction between the electrons has to be taken into account,
even for a theory of the DOS in the region of the IQHE which is commonly
believed to be explainable without electron-electron interactions.
More quantitative comparisons between experiment and
theory are necessary, in order to clarify this point.

If our above suspicion proves to be true, the proper understanding
of the IQHE would very probably need, in addition to disorder,
electron-electron interaction as an important ingredient, as
is already the case for the fractional quantum Hall effect.

\smallskip
We thank Maura Sassetti and Wolfgang Hansen for useful discussions
and the latter also  for providing the recent experimental data
prior to publication \cite{h96}.
This work was supported by the
Deut\-sche For\-schungs\-ge\-mein\-schaft
within the Graduiertenkolleg ''Physik nanostrukturierter Festk\"orper''
of the Universit\"at Hamburg and via projects Ap 47/1-2 and kr627/8,
and by the EU via contracts CHRX-CT93 0136 and FMRX-CT96 0042.

%%%%%%%%%%%%%%%%%%%%%%%%%%%%%%%%%%%%%%%%%%%%%%%%%%%%%%%%%%%%%%%%

\begin{figure}[p]
\caption[erg] {\label{erg}
DOS, in units of the DOS at $B = 0$, for the eight lowest LLs
for $L=6$ and different
strengths of disorder: thin solid line $\eta = 0.2$, dotted line
$\eta = 0.3$, thick solid line $\eta = 0.4$.}
\end{figure}

%%%%%%%%%%%%%%%%%%%%%%%%%%%%%%%%%%%%%%%%%%%%%%%%%%%%%%%%%%%%%%%

\begin{figure}[p]
\caption[bhl] {\label{bhl}
Thick solid line: DOS of the present model; thin solid line:
approximation for a model with Gaussian correlations
\protect\cite{brod} for
$L = 10$ and $\eta = 0.4$.}
\end{figure}

%%%%%%%%%%%%%%%%%%%%%%%%%%%%%%%%%%%%%%%%%%%%%%%%%%%%%%%%%%%%%%%%

\begin{figure}[p]
\caption[exp] {\label{exp}
Normalized DOS at the cyclotron energy
versus the magnetic field strength. Solid line: experiment
of  \protect\cite{eetal85};
solid line with bullets: recent experiment
 of \protect\cite{h96}.
Results for the present model:
$\eta = 0.67\;\mbox{meV}$; $L = 1089\cdot 10^{-8}\mbox{cm}\,(\triangle )$,
$\eta = 1.26\;\mbox{meV},\, L = 787\cdot 10^{-8}\mbox{cm}\,(\bigcirc )$,
$\eta = 1.86\;\mbox{meV},\, L = 485\cdot 10^{-8}\mbox{cm}\, (\Diamond )$.}
\end{figure}

%%%%%%%%%%%%%%%%%%%%%%%%%%%%%%%%%%%%%%%%%%%%%%%%%%%%%%%%%%%%%%%%%%

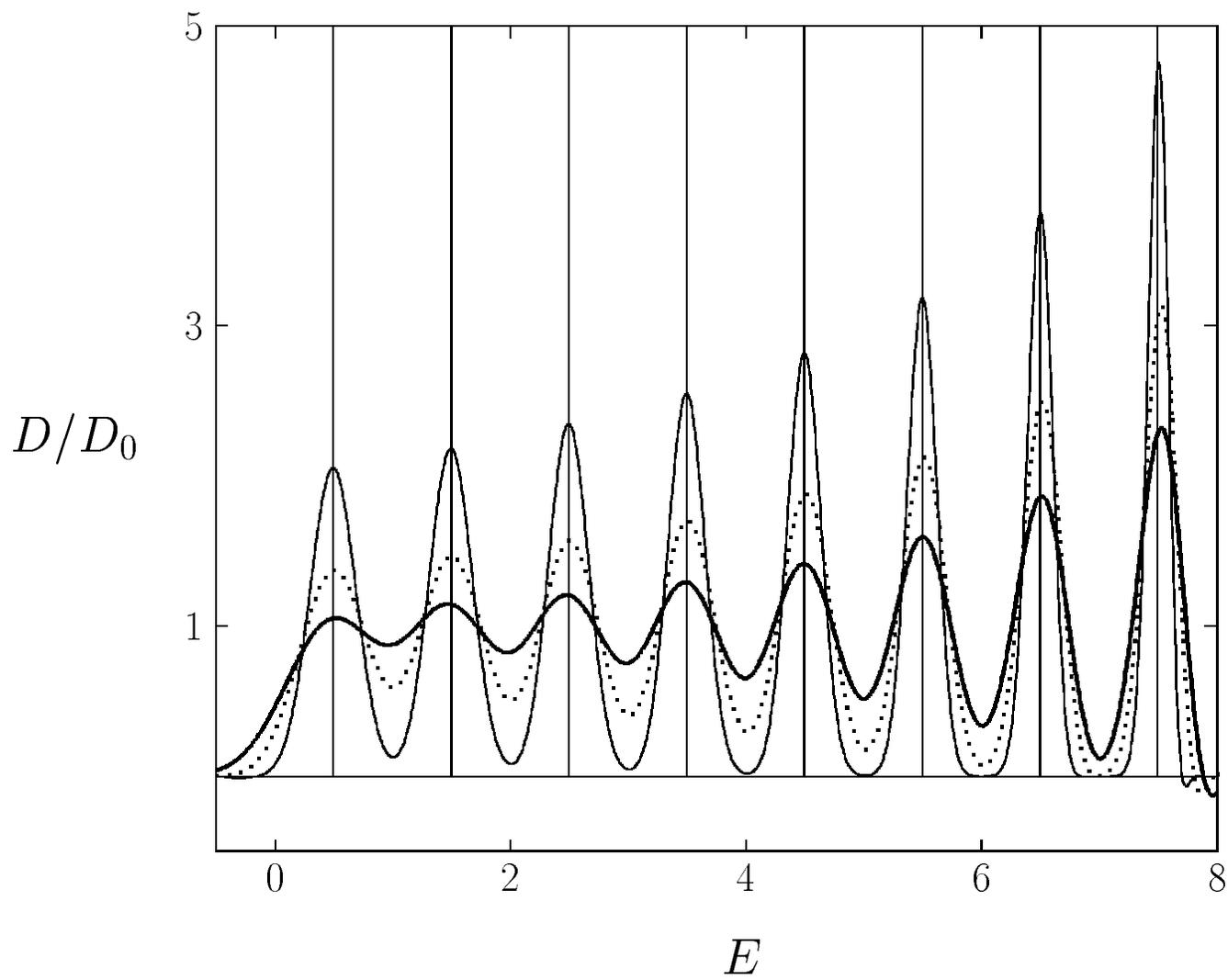
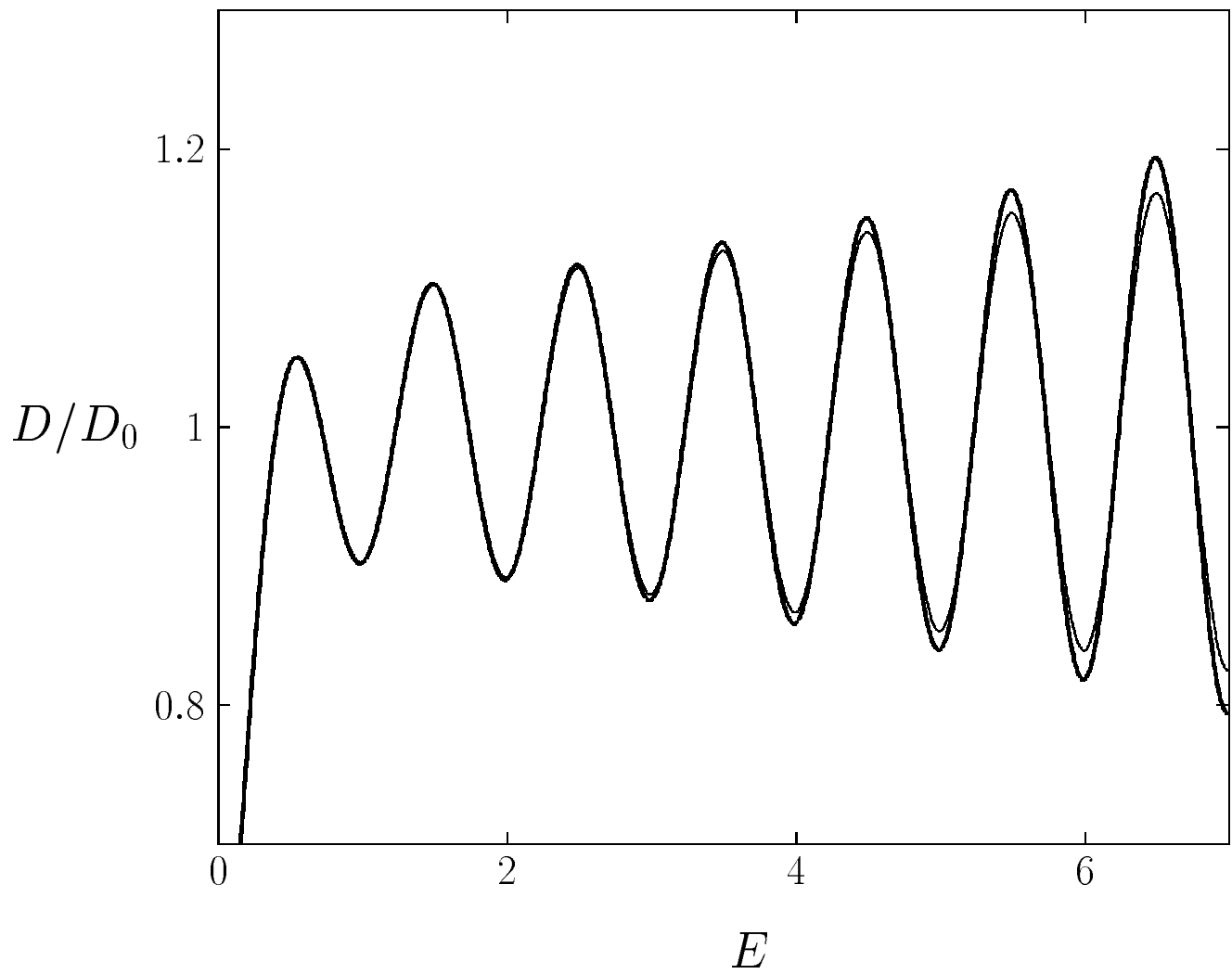
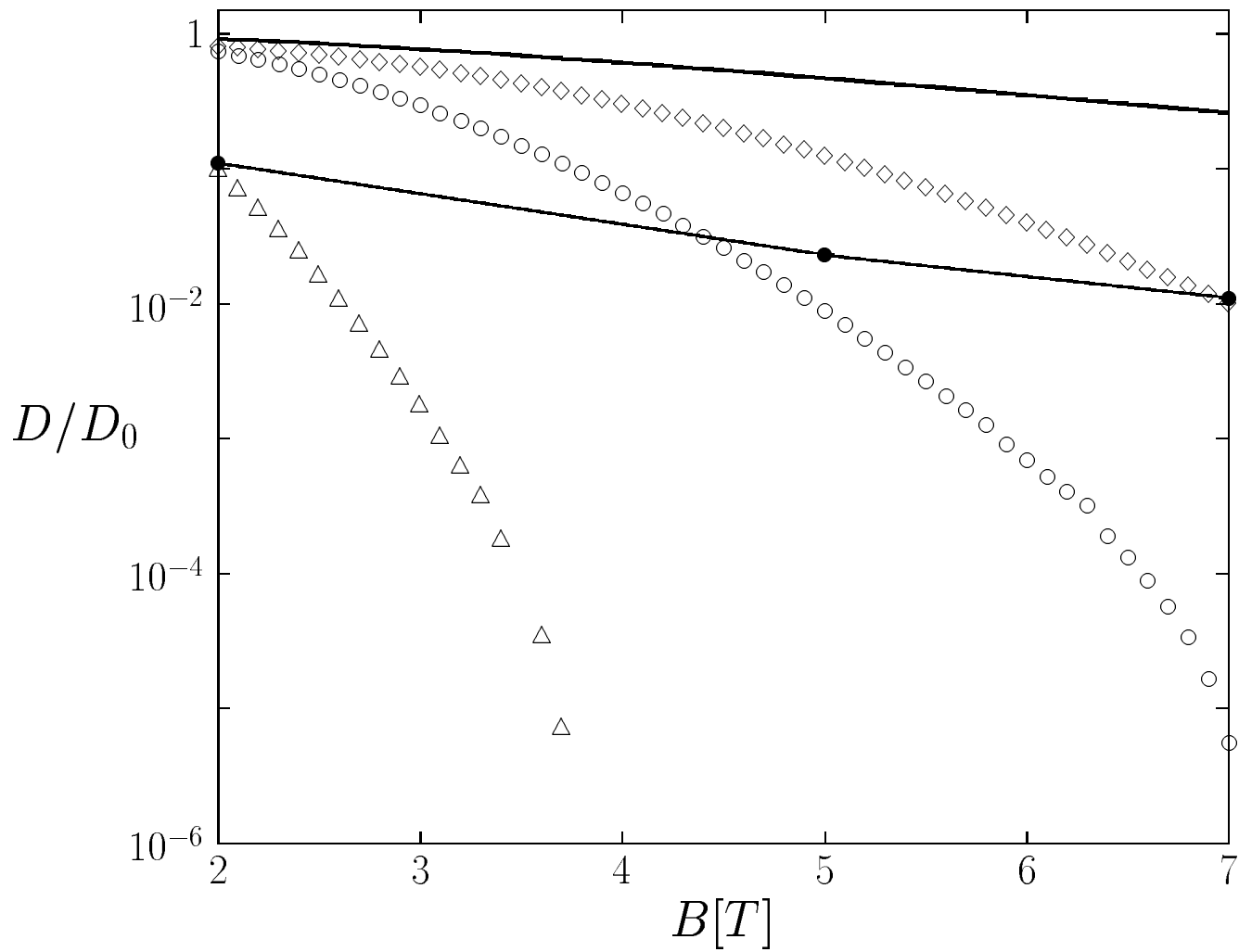
\end{document}